\documentclass[preprint,showpacs,preprintnumbers,amsmath,amssymb]{revtex4}
\usepackage{amsmath}
\usepackage{graphicx}
\usepackage{amssymb}
\usepackage{bm}
\begin{document}

\title{Competition between fusion-fission and quasifission processes in the $^{32}$S+$^{182,184}$W reactions }
\author{H. Q. Zhang}
\altaffiliation{School of Physics, Peking University 100871 Beijing, China}
\email{huan@ciae.ac.cn}
\author{C. L. Zhang}
\affiliation{China Institute of Atomic Energy, P. O. Box 275, Beijing 102413, China}
\author{C. J. Lin}
\affiliation{China Institute of Atomic Energy, P. O. Box 275, Beijing 102413, China}
\author{Z. H. Liu}
\affiliation{China Institute of Atomic Energy, P. O. Box 275, Beijing 102413, China}
\author{F. Yang}
\affiliation{China Institute of Atomic Energy, P. O. Box 275, Beijing 102413, China}
\author{A. K. Nasirov}
\altaffiliation{Institute of Nuclear Physics, 100214, Tashkent, Uzbekistan}
\email{nasirov@jinr.ru}
\affiliation{Joint Institute for Nuclear Research, 141980 Dubna, Russia }
\author{G. Mandaglio, M. Manganaro, and G. Giardina}
\affiliation{Dipartimento di Fisica dell' Universit\`a di Messina, 98166 Messina,  and Istituto Nazionale di Fisica Nucleare, Sezione di Catania,  Italy}

\begin{abstract}
The angular distributions of fission fragments for the $^{32}$S+$^{184}$W
reaction at center-of-mass energies of 118.8, 123.1, 127.3, 131.5, 135.8, 141.1
and 144.4 MeV were measured. The experimental fission excitation function is obtained. The fragment angular anisotropy ($\mathcal{A}_{\rm exp}$) is found by extrapolating the each fission angular distributions. The measured fission cross sections of the $^{32}$S+$^{182,184}$W reaction are decomposed into fusion-fission, quasifission and fast fission contributions by the dinuclear system model. The total evaporation residue excitation function for  the $^{32}$S+$^{184}$W reaction calculated in the framework of the advanced statistical model is in good agreement with the available experimental data up to about $E_{\rm c.m.}\approx 160$ MeV. The theoretical descriptions of the experimental capture excitation functions for both reactions and quantities $K_0^2$,  $<\ell^2>$ and  $\mathcal{A}_{\rm exp}$ which characterize angular distributions of the fission products were performed by the same partial capture cross sections at the considered range of beam energy.
\end{abstract}

\date{Today}
\maketitle

\section{Introduction}
\label{sec:1}

Studies of fusion-fission reactions between heavy ion
projectile and heavy target nuclei have demonstrated to be very useful in developing an understanding of the nuclear reaction dynamics. Especially with the development of radioactive nuclear beams and the superheavy element
synthesis, this study is becoming a hot topic again. Very recently, the
synthesis of  the heaviest elements of 114, 115, 116 and 118  by using the hot-fusion reactions \cite{Oganessian04,FLNR} with actinide targets and of 112 and 113  by using the cold-fusion reactions \cite{Morita73,Morita76} with lead-based targets of shell closed spherical nuclei have been reported. The cross section of the evaporation residue (ER) formation being a superheavy element is very small: some picobarns, or even some percents of picobarn at synthesis of the element $Z$=113.

There are two main reasons causing a hindrance to the ER formation
in the reactions with massive nuclei: the quasifission and
fusion-fission processes. The ER formation process is often
considered as third stage of the three-stage process. The first
stage is a capture--formation of the dinuclear system (DNS) after
full momentum transfer of the relative motion of colliding nuclei
into the deformed shape, excitation energy and rotational energy.
The capture takes place if the initial energy of projectile in the
center-of-mass system is enough to overcome the interaction
barrier (Coulomb barrier + rotational energy of the entrance channel).
The study of dynamics of processes in heavy ion collisions at the near Coulomb barrier energies showed that complete fusion
does not occurs immediately in the case of the massive nuclei
collisions \cite{Back32,VolPLB1995,AdamianPRC2003,FazioEPJ2004}.
The quasifission process competes with formation of compound
nucleus (CN). This  process occurs when the DNS 
prefers to break up into fragments instead of to be transformed
into fully equilibrated  CN. The number of events going to
quasifission increases drastically by increasing  the sum of the
Coulomb interaction and rotational energy in the entrance channel
\cite{GiaEur2000,FazioPRC2005}. Another reason decreasing yield of
ER is the fission of a heated and rotating CN which is formed in
competition with quasifission. The stability of massive CN
decreases due to the decrease in the fission barrier by increasing
its excitation energy $E^*_{\rm CN}$ and angular momentum $L$
\cite{ArrigoPRC1992,ArrigoPRC1994,SagJPG1998}. Because the stability of the
transfermium nuclei are connected with the availability of shell
correction in their binding energy \cite{Sobiczewski,Muntian}
which are sensitive to  $E^*_{\rm CN}$ and  values of the angular
momentum. To find favorable reactions (projectile and target pair)
and the optimal beam  energy range leading to larger cross
sections of synthesis of superheavy elements, we should establish
conditions to increase as possible the events of ER formation.

The total evaporation residue and fusion-fission excitation functions for  the $^{32}$S+$^{182,184}$W reactions are  calculated in the framework of the advanced statistical model \cite{ArrigoPRC1992,ArrigoPRC1994,SagJPG1998}. The results of calculation are in good agreement with the experimental data presented in Ref. \cite{Back60} for the $^{32}$S+$^{184}$W reaction up to $E_{\rm c.m.}\simeq 160$ MeV. The dip of the theoretical curve from experimental data  at high excitation energies $E_{\rm c.m.}\simeq 160$ MeV is caused by the fact that statistical model can not reproduce the cross section of formation of reaction products by the nonequlibrium  mechanism without formation of the compound nucleus in equlibrium state.  

To determinate the ER cross section $\sigma _{\rm ER}(E)$ we used the partial fusion cross section as initial data about the heated and rotating CN with given excitation energy $E$ and angular momentum $\ell$ \cite{VolPLB1995}:  
\begin{equation}
\label{ER}
\sigma _{\rm ER}(E)=\sum\limits_{l=0}^{\infty }(2l+1)\sigma^{(l)}_{\rm fus}(E)W_{\rm sur}(E,l).
\end{equation}%
The entrance channel effects can be studied \cite{NasirovNPA759} by analyzing the partial fusion cross section $\sigma^{l}_{\rm fus}(E)$ which is defined by the expression:
\begin{equation}
\label{fusion}
\sigma^{(l)}_{\rm fus}(E)=\sigma^{(l)}_{\rm capture}(E)P_{\rm CN}(E,l).
\end{equation}
 The theoretical cross section of capture includes the
contributions of all fragment yields from full momentum transfer
reactions:
\begin{equation}
\label{sumcap}
   \sigma_{\rm cap}(E_{\rm c.m.})=
   \sigma_{\rm ER}(E_{\rm c.m.})+
   \sigma_{\rm f}(E_{\rm c.m.})+
   \sigma_{\rm qf}(E_{\rm c.m.})+
   \sigma_{\rm fast\,fission}(E_{\rm c.m.}),
\end{equation}
where $\sigma_{\rm ER}$, $\sigma_{\rm f}$,  $\sigma_{\rm qf}$,  and
$\sigma_{\rm fast\,fission}$ are  the evaporation residue, fusion-fission, quasifission and fast fission cross sections, respectively.

The pure cross section of the complete fusion must include only evaporation residues and fusion-fission cross sections  
\begin{equation}
\sigma^{(\rm pure)}_{\rm fus}=\sigma_{\rm ER}+\sigma_{\rm ff}.
\end{equation}

The  experimental value of $\sigma_{\rm fus}$ reconstructed from the detected  fissionlike fragments and evaporation residues:
\begin{equation}
\sigma_{\rm fus}=\sigma_{\rm ff}+
\sigma_{\rm qf}+\sigma_{\rm fast\,fis}+\sigma_{\rm ER},
\end{equation}
where $\sigma_{\rm ff}$, $\sigma_{\rm qf}$, and
$\sigma_{\rm fast\,fis}$  are the contributions of fusion-fission, quasifission and  fast fission processes, respectively, and $\sigma_{\rm ER}$ is the ER contribution.

Thus, the estimations of $P_{\rm CN}$ and $W_{\rm sur}$ are the key point for the research of the fusion reaction products, especially for the synthesis of superheavy elements.

The presence of quasifission fragments in the measured yield of
fissionlike fragments is determined by the large values of
anisotropy in their angular distribution (see
Refs.\cite{Back32,HindePCR53,anisEPJA34}, and references therein)
and by the increasing yield of fragments with masses near  proton
magic numbers 28, 50, 82 and neutron magic numbers 50, 82, 126
\cite{Itkis,KnyazhevaPRC75}. But the  mass
and angular distributions of quasifission products can overlap with ones of fusion-fission products. The sizes of each overlap in the both  mass and angular
distributions  depend on the beam energy and mass asymmetry of
reacting nuclei. Therefore, quantitative estimations of
contributions of quasifission and fusion-fission fragments in the
measured fissionlike fragments is a nontrivial task up to now.
This means that the estimation of the pure fusion cross section
from the measured data of fission fragments and evaporation
residues  is still an ambiguous task.

In this paper we have analyzed  the angular distribution of
fission fragments and fission excitation function of the $^{32}$S+$^{184}$W reaction measured at the center-of-mass energies of 118.8, 123.1, 127.3, 131.5, 135.8, 141.1 and 144.4 MeV in this work and similar data obtained from Ref. \cite{Keller} at more large energies $E_{\rm c.m.}=141.2$--221.1 MeV. 
The angular anisotropy $\mathcal{A}_{\rm exp}$ of fission fragments 
measured in this work was found by extrapolating the each fission angular distributions to angles 0$^{\circ}$ and 90$^{\circ}$  by the method used in Ref.
\cite{Hui177}.  Then, the mean square angular momentum $\langle
{L^{2}}\rangle$ values were obtained. Hereafter we use for
simplicity $\ell$ from the definition $L=\ell \hbar$. 

  The experimental data of the quantities $K_0^2$ and anisotropy of the angular distribution of the fission products of the $^{32}$S+$^{182}$W \cite{Keller} and $^{32}$S+$^{184}$W (this work) reactions are described using angular momentum distribution of DNS calculated as a function of the orientation angle of symmetry axis of $^{182}$W and $^{184}$W.

We assumed the calculated  capture cross sections to be equal to the
experimental data of fissionlike fragments in order to decompose
the measured fission cross section into fusion-fission,
quasifission and fast fission contributions by the DNS model \cite{VolPLB1995,FazioEPJ2004,FazioPRC2005,NasirovNPA759}. We
remind the difference between  quasifission and fast fission. The
quasifission is a break of the DNS into two fragments bypassing
the stage of the CN formation. The fast fission process is the
inevitable decay of the fast rotating mononucleus into two
fragments without reaching the equilibrium compact shape of CN.
Such mononucleus is formed from the DNS survived
against quasifission. At large values of the angular momentum
$\ell > \ell_f$, where $\ell_f$ is a value of $\ell$ at which the
fission barrier of the corresponding CN disappears,  mononucleus
immediately decays into two fragments  \cite{Gregoire}. As
distinct from fast fission, the quasifission can occur at all
values of $\ell$ at which capture occurs. 

The present article is organized in the following way. The experimental
procedure is presented in Section~\ref{sec:2}. The experimental results
of the fission fragment angular distributions and their anisotropy  $A_{\rm exp}$, fission excitation function,  mean square value   $\langle {l^{2}}\rangle $ of angular momentum and variance $K_0^2$  of the $K$ distribution are presented in  Section~\ref{sec:3}. The comparison between the experimental data and  theoretical results is discussed in Section~\ref{sec:4}. Finally, Section~\ref{sec:5} is devoted to the conclusion of this work.

\section{EXPERIMENTAL PROCEDURE}
\label{sec:2}

The experiment was performed at HI-13 tandem accelerator at China
Institute of Atomic Energy, Beijing. A collimated $^{32}$S beam
with incident energies $E_{\rm lab}=140,145,150,155,160,165$ and
$170$ MeV bombarded on a target of $^{184}$W which was mounted at
center of the scattering chamber. The $^{184}$W target with
thickness about 200 $\mu $g/cm$^{2}$ was evaporated on an about 20
$\mu $g/cm$^{2}$ carbon foil backing. Typical $^{32}$S beam
current range was 800-1000 enA monitored by a shielded suppressed
Faraday cup at periphery of the chamber, because of the variety
according to the bombarding energy and scattering angle. The beam
energy loss in traveling half  the target were calculated and was
about 0.5 MeV.
\begin{figure}
\vspace*{-2.0cm}
\begin{center}
\resizebox{0.80\textwidth}{!}{\includegraphics{{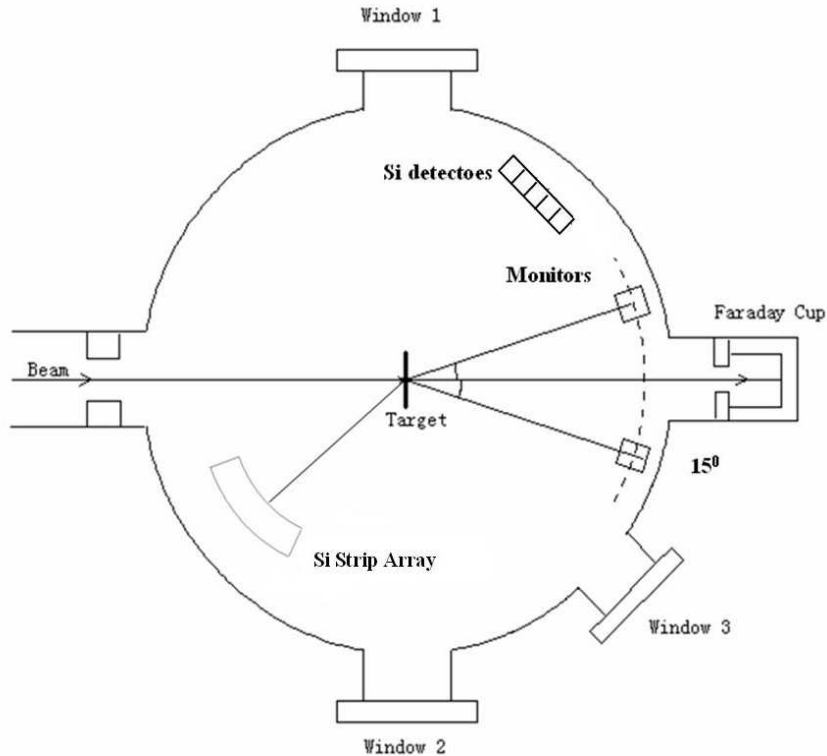}}}
\vspace*{-1.35 cm} \caption{\label{fig_1} Scheme of the experimental
setup for the measurements of fission fragment angular
distribution of binary reaction products. See explanations in the
text for details.}
\end{center}
\end{figure}

The schematic view of the experimental set up is shown in Fig.~\ref{fig_1}.
At the forward angles, an array of five Si detectors with depletion depth
ranging from 200 to 300 $\mu $m, which covered the angular range of
$\theta_{L}$=14$^{\circ}$-35$^{\circ}$, 35$^{\circ}$-55$^{\circ}$ and 55$^{\circ}$-75$^{\circ}$ were mounted on the movable arm in the chamber and five masks were placed in the front of each detector for assuring the angular resolution. The detectors to the target distance was 27 cm. The Rutherford scattering was monitored at forward angle of $\theta _{L}$=15$^{\circ}$ by four Si(Au) surface barrier detectors for the normalization of the cross section measurements.

In addition to these individual Si detectors, two groups of Si strip
detectors were mounted on opposite sides of the beam. These Si strip
detectors were $48\times 50$ mm$^2$ in area and each detector consisted of 24
strips. Due to a lack of readout electronics, the strips were tied together
in groups of eight for readout. Data from these strip detectors were
recorded in the coincidence mode with the requirement that each detector was
struck by a fission fragment and the folding angle between the hits
corresponded to a full momentum transfer event. To calculate the kinematics
it was assumed all observed  processes can be treated as binary reactions. This
assumption was tested by examining the folding angle distribution of
coincident fragments in the Si detector and the Si strip detectors. The
average folding angle agrees with the expectations based on total fission
kinetic energies taken from the Viola systematics \cite{Viola1985}.

To obtain the absolute cross sections, we  measured the solid angle by
using the $\alpha $-particles from the $^{241}$Am source and the elastic
scattering products. Centering of the beam on the target was ensured by the
four monitor detectors. A gate was set on the fission event and data were
collected by the coincident mode.

\begin{figure}
\vspace*{-2.0cm}
\begin{center}
\resizebox{0.80\textwidth}{!}{\includegraphics{{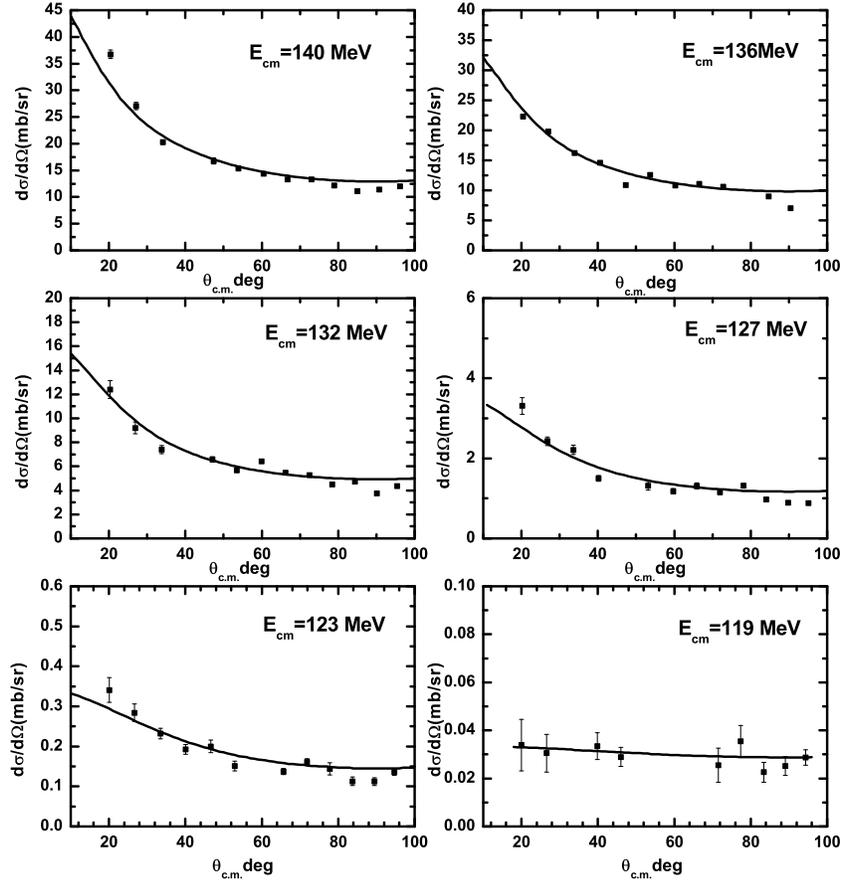}}}
\vspace*{-1.35 cm} \caption{\label{fig_2} Fission fragment angular
distributions for the  $^{32}$S+$^{184}$W reaction. Incident
energies are shown in the figure. The experimental data are shown
with the fitting curve, which is used to determine the anisotropy
$A_{\rm exp}$ of the fragment angular distribution and mean square
values of angular momentum  from these events.}
\end{center}
\end{figure}
\section{EXPERIMENTAL RESULTS}

\label{sec:3}

\subsection{Fission fragment angular distributions}

The fission fragment angular distributions were measured using the
coincident detectors and are shown in Fig.~\ref{fig_2}.
In fitting the angular distribution  of the fission fragments we
used the familiar expression as in Ref. \cite{Hui177}:
\begin{equation}
\label{Wangl}
W(\theta )=\sum\limits_{J=0}^{J_{\rm max}}{\frac{{(2J+1)^{2}\exp[-(J+{\frac{1}{2}}%
)^{2}\sin^{2}\theta /4K_{0}^{2}]J_{0}[i(J+{\frac{{1}}{{2}}}%
)^{2}\sin^{2}\theta /4K_{0}^{2}]}}{{{\rm erf}[(J+{\frac{{1}}{{2}}}%
)/(2K_{0}^{2})^{1/2}]}}}
\end{equation}%
assuming $M$=0, {\it i.e.} assuming the spins of the target and projectile were
zero, where $J_{0}$ is the zero order Bessel function with imaginary argument and error function ${\rm erf}(x)$ is defined as
\begin{equation}
{\rm erf}(x)=(2/\pi ^{1/2})\int\limits_{0}^{x}\exp(-t^{2})dt.
\end{equation}
The measured values of $\sigma_{\rm capture}$ and  the deduced values
of $\mathcal{A}_{\rm exp}$ and $K^2_0$ for the $^{32}$S+$^{184}$W reaction
are presented in Table \ref{table1}.  $J_{\rm max}$ is obtained by reproducing the capture cross section. The $K^2_{0}$ value  is found by fitting the angular distribution at known $J_{\rm max}$ from the total fission cross section.
It is seen from Fig. \ref{fig_2} that the anisotropy of angular distribution increases by increasing collision energy $E_{\rm c.m.}$.

\medskip \medskip \medskip \medskip \thinspace \thinspace \thinspace
\begin{table}[tbp]
\caption{ The measured capture cross sections and the deduced values
of $\mathcal{A}_{\rm exp}$ and $K^2_0$ for the $^{32}$S+$^{184}$W reaction. The $E_{\rm c.m.}$  are the energies calculated as corresponding to the beam energies in the center of the target.}
\label{table1}\centering
\begin{tabular}[t]{ccccccccccc}
\hline\hline
& $E_{\rm c.m.}$ &  & $E^{*}_{\rm CN}$ &  & $\sigma_{\rm capture}$ &  &  $ \mathcal{A}_{\rm exp} $ &   &  $ K^2_{0} $  \\
& (MeV) &  & (MeV)&  & (mb)   &  &       & & \\ \hline
& 118.8 &  & 37.2 &  & 0.04   &  &  1.51 & &  114.71 \\
& 123.1 &  & 41.5 &  & 2.35   &  &  2.16 & & 124.35  \\
& 127.3 &  & 45.8 &  & 22.97  &  & 2.27  & & 132.09\\
& 131.5 &  & 50.0 &  & 81.01  &  &  2.74 & & 140.01 \\
& 135.8 &  & 54.3 &  & 132.27 &  & 3.06  & & 148.67 \\
& 141.1 &  & 58.5 &  & 189.33 &  & 3.28  & & 157.35 \\
& 144.4 &  & 61.8 &  & 237.06 &  & 3.80  & &  155.09\\ \hline\hline
\end{tabular}
\thinspace
\end{table}

\subsection{Capture cross section}

In order to deduce the capture cross section from the data, the Si
strip detectors  operating in the coincidence mode were used. Each
fission event was selected on the basis of the correct value of
energy and of the folding angle corresponding to complete momentum
transfer using the forward Si detectors as a "trigger" detector.
After correction for the efficiency of the Si strip detectors, a
differential cross section $d\sigma /d\Omega (\theta )$ was
obtained. The total cross section was deduced from the integration
of the differential cross sections.

The resulting experimental  values  of the capture cross sections
are shown in Table \ref{table1} and Fig.~\ref{CrossSec84} where they
are compared with the theoretical results. Due to the small cross section
of ER in the reaction under consideration \cite{Back60}, the total measured
fission cross section is assumed to be equal to the theoretical
capture cross section and it was decomposed into  fusion-fission,
quasifission and fast fission parts in the framework of the DNS
model mentioned in Section \ref{sec:1}.
The comparison of the measured capture cross
section, anisotropy, mean square values of angular momentum, and variance $K_0^2$  with the corresponding experimental data is discussed in Section \ref{sec:4}.

\subsection{Anisotropy of fission-fragments and mean square angular momentum
values}

The experimental values of the anisotropy $\mathcal{A}_{\rm exp}$ are found  by extrapolating the fission angular distributions to angles 0$^{\circ}$ and 90$^{\circ}$  by the method used in Ref. \cite{Hui177}.
The anisotropies as a function of center-of-mass energies are shown in
Table \ref{table1} and Fig. \ref{FigAnis} where they are compared with the theoretical results. Using the approximate relation between the anisotropy and the mean square angular momentum (see Section~\ref{sec:4} for details), the mean square angular momentum values $\langle l^{2}\rangle $ are
deduced from the experimental anisotropies and shown in Fig. ~\ref{FigL2}.

\section{Theoretical description  and comparison with measured data}

\label{sec:4}

The experimental data for the excitation function of fissionlike products
in the $^{32}$S+$^{184}$W reaction were analyzed in the framework of the
DNS model \cite{GiaEur2000,FazioPRC2005,NasirovNPA759,NasirovPRC79,FazioJSot2008}. 
The capture, fusion, quasifission, fusion-fission, evaporation residue, and fast fission excitation functions have been calculated for this reaction.
\begin{figure}[tbp]
\vspace*{1.5cm}
\par
\begin{center}
\resizebox{0.60\textwidth}{!}{\includegraphics{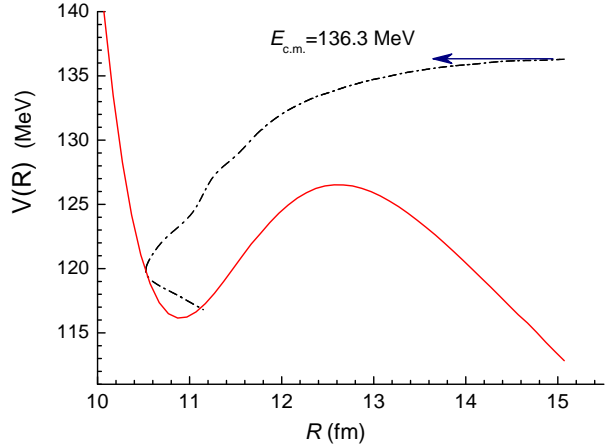}}
\vspace*{-3.7cm}
\end{center}
\caption{(Color online) Illustration of capture path (dot dashed line) into potential well
(solid line) as the numerical solution of the equation of relative motion of
colliding nuclei with the initial energy $E_{\rm c.m.}=136.3$ MeV and $L$=0 for
the $^{32}$S+$^{184}$W reaction.}
\label{capture}
\end{figure}
According to the DNS model a capture event is the trapping of the collision
path into the potential well  (see Fig. \ref{capture}) after dissipation of the sufficient part of the relative kinetic energy of a projectile nucleus in the center-of-mass  coordinate system. At capture the full momentum
transfer from the relative motion of nuclei into excitation energy of
dinuclear system takes place. Certainly the presence of a potential pocket
and adequacy of the collision energy $E_{\rm c.m.}$ to overcome  the interaction barrier of the entrance channel $V_{\rm B}$ are necessary conditions to occur capture as shown in Fig. \ref{capture}. Thus capture leads to forming DNS which characterized by mass (charge) asymmetry of its nuclei, rotational energy $V_{\rm rot}$  and excitation energy $E^*_{\rm DNS}$. The relative energy of nuclei is relaxed, therefore, the total kinetic energy of fragments formed at its decay are close to the Viola systematics \cite{Viola1985}.

The nucleus-nucleus potential $V(Z,A,R)$ is a sum of the Coulomb
$V_{\rm C}(Z,A, R)$ and nuclear interaction $V_{\rm N}(Z,A,R)$, as well as the rotational energy $V_{\rm rot}(Z,A,R,\ell)$:
\begin{equation}
V(Z,A,R,\ell)=V_{\rm C}(Z,A,R)+V_{\rm N}(Z,A,R)+V_{\rm rot}(Z,A, R,\ell)
\end{equation}
where $Z=Z_1$ and $A=A_1$ are the charge and mass numbers of one of fragments forming the DNS, respectively,  while the charge and mass numbers of another fragment are equal to  $Z_2=Z_{\rm CN}-Z$ and $A_2=A_{\rm CN}-A$, respectively, where  $Z_{\rm CN}$ and $A_{\rm CN}$ the charge and mass numbers of being formed CN; $R$ is the relative distance between the centers of nuclei forming DNS.
\begin{figure*}
\vspace{-0.5 cm}
\begin{center}
\resizebox{0.80\textwidth}{!}{\includegraphics{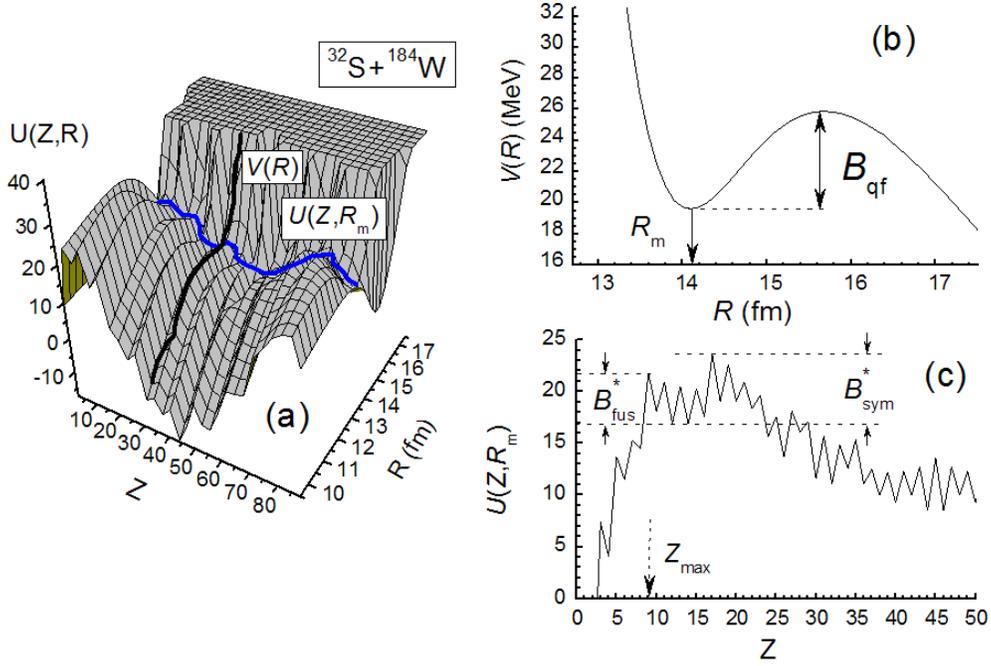}}
\vspace{-0.5cm} \caption{\label{PES216Th} (Color online) The potential energy
surface for a dinuclear system leading to the formation of the
$^{216}$Th$^*$ CN   as a function of the relative distance $R$
between centers of interacting nuclei and their charge numbers
$Z$, panel (a);  the nucleus-nucleus interaction potential $V(R)$
shifted on the $Q_{gg}$-value for the $^{32}$S + $^{184}$W
reaction, panel (b); the driving potential, $U_{\rm dr}(Z,R_{m})$,
which is a curve linking minimums corresponding to each charge
asymmetry in the valley of the potential energy surface as a
function of  $Z$, panel (c).} \vspace*{10pt}
\end{center}
\end{figure*}
The partial capture cross section is found by solution of kinetic equation
for the relative motion and orbital angular momentum $\ell$ for the different
orientation angle $\alpha_{\rm T}$ of the target nucleus as it was performed in Refs. \cite{GiaEur2000, NasirovNPA759}. The fusion cross section is calculated from the branching ratio $P_{\rm CN}(Z)$ of the decay rates of overflowing the border of the potential well ($B^{(Z)}_{\rm qf}$) along $R$ at a given mass asymmetry (decay of DNS--quasifission) over the barriers on mass asymmetry axis $B^{*}_{\rm fus}$ for the complete fusion or $B^{(Z)}_{\rm sym}$ in opposite direction to the symmetric configuration of DNS (see Fig. \ref{PES216Th}):
\begin{equation}
\label{PcnG}
P^{(Z)}_{\rm CN}(E^*_{\rm DNS})\approx{\frac{{\Gamma^{(Z)}_{\rm fus}}(B^{*}_{\rm fus},E^*_{\rm DNS})}{{\Gamma^{(Z)}_{(\rm qf)}(B_{\rm qf},E^*_{\rm DNS})+\Gamma^{(Z)}_{(\rm fus)}(B^{*}_{\rm fus},E^*_{\rm DNS})+\Gamma^{(Z)}_{\rm sym}(B_{\rm sym},E^*_{\rm DNS})}}},
\end{equation}
where $\Gamma_{\rm fus}$,  $\Gamma_{\rm qf}$ and  $\Gamma_{\rm sym}$ are corresponding widths determined by the level densities on the barriers  $B^{*}_{\rm fus}$,  $B_{\rm qf}$ and $B^{*}_{\rm sym}$  involved in the calculation of $P_{\rm CN}$ are used in the model \cite{NasirovNPA759, FazioPRC2005, AdamianPRC2003} based on the DNS concept \cite{Volkov}.  Here $E^{*}_{\rm DNS}(Z_{\rm P},A_{\rm P},\ell)= E_{\rm c.m.}-
V(Z_{\rm P},A_{\rm P},\ell,R_{\rm m})$ is the excitation energy of DNS in the entrance channel, where $Z_{\rm P}$ and $A_{\rm P}$ are charge and mass numbers of the projectile nucleus. $V(Z,A,R_{\rm m},\ell)$ is the minimum value of the nucleus-nucleus potential well (for the DNS with charge asymmetry $Z$) and  its position on the relative distance between the centers of nuclei is marked as $R=R_{\rm m}$ in Fig.  \ref{PES216Th}b. The value of $B_{\rm qf}$ for the decay of DNS with the given charge asymmetry of fragments is equal to the depth of the potential well in the nuclear-nuclear  interaction (see Fig. \ref{PES216Th}b). The intrinsic fusion barrier  $B^{*}_{\rm fus}$ is connected with mass (charge) asymmetry degree of freedom of the DNS and it is determined from the potential energy surface (Fig. \ref{PES216Th}a):
\begin{equation}
U(Z;R,\ell)=U(Z,\ell,\beta_{1},\alpha_{1};\beta_{2},\alpha_{2}) =B_{1}
+B_{2}+V(Z,\ell,\beta_{1},\alpha_{1};\beta_{2},\alpha_{2};R)-
(B_{\rm CN}+V_{\rm CN}(\ell)).
\end{equation}
Here, $B_{1}$, $B_{2}$ and $B_{\rm CN}$ are the binding energies of the nuclei in DNS and the CN, respectively, which were obtained from \cite{MollerADND1988}; the fragment deformation parameters $\beta _{i}$ are taken from the tables in \cite{RamanADND1987, SpearADND1989, MollerADND1988} and $%
\alpha _{i}$ are the orientation angles of the reacting nuclei
relative to the beam direction; $V_{\rm CN}(\ell)$ is the
rotational energy of the CN. The distribution of neutrons between
two fragments for the given proton numbers $Z$ and $Z_{2}$ or
ratios $A/Z$ and $A_{2}/Z_{2}$ for both fragments were determined by minimizing the potential $U(Z;R)$ as a function of $A$ for each $Z$.

The driving potential $U_{\rm dr}(Z)\equiv U(Z,R_{\rm m})$ is a curve linking minimums corresponding to each charge asymmetry $Z$ in the valley of the potential energy surface from $Z=0$ up to $Z=Z_{\rm CN}$ (see  Fig. \ref{PES216Th}a and \ref{PES216Th}c).  We define the intrinsic fusion barrier for the DNS with charge asymmetry $Z$
as $B_{\rm fus}^{\ast }(Z,\ell)=
U(Z_{\rm max},R_{\rm m}(Z_{\rm max}),\ell)-U(Z,R_{\rm m}(Z),\ell)$, where $U(Z_{\rm max},\ell)$ is a maximum value of potential energy at  $Z=Z_{\rm max}$ in the valley along the way of complete fusion from the given $Z$ configuration. The $B_{\rm sym}^{\ast }(Z,\ell)$ is defined by the similar way as shown in Fig. \ref{PES216Th}c \cite{NasirovNPA759,FazioMPL2005}.

The masses and charges of the projectile and target nuclei are not constant during capture and after formation of the DNS. The intense proton and neutron exchange between constituents of DNS is taken into account by
calculation of the complete fusion probability $P_{\rm CN}$ as fusion from all
populated DNS configurations according to the formula
\begin{equation}
\label{PcnY}
P_{\rm CN}(E^{*}_{\rm DNS}(Z,A,\ell);\{\alpha_{i}\})=\sum\limits_{Z_{\rm sym}}^{Z_{\rm max}}Y_Z(E^{*}_{\rm DNS}(Z,A,\ell))P^{(Z)}_{\rm CN}(E^{*}_{\rm DNS}(Z,A,\ell);\{\alpha_{i}\})
\end{equation}
where $E^{*}_{\rm DNS}(Z,A,\ell)= E^{*}_{\rm DNS}(Z_{\rm P},A_{\rm P},\ell)+\Delta{Q_{\rm gg}(Z)}$ is the excitation energy of DNS with angular momentum $\ell$ for a given value of its charge-asymmetry configuration $Z$ and $Z_{\rm CN}-Z$;
$Z_{\rm sym}=(Z_1+Z_2)/2$;  $\Delta{Q_{\rm gg}}(Z)$ is the change of $Q_{\rm gg}$-value by changing the charge (mass) asymmetry of DNS; $Y_Z(E^{*(Z)}_{\rm DNS})$ is the probability of population of the ($Z, Z_{\rm CN}$-Z) configuration  at $E^{*(Z)}_{\rm DNS}$ and given orientation angles ($\alpha_1,\alpha_2$). It
was obtained by solving  the master equation for the evolution of the
dinuclear system charge asymmetry (for details see Refs. \cite%
{NasirovNPA759,anisEPJA34}).

\begin{figure}[t]
\vspace*{-0.40cm}
\par
\begin{center}
\resizebox{0.80\textwidth}{!}{\includegraphics{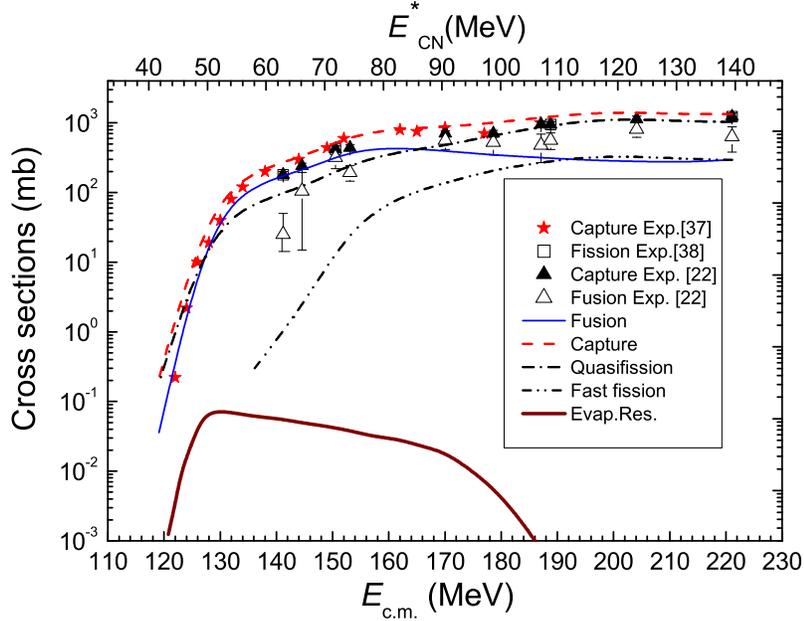}}
\vspace*{-1.6cm}
\end{center}
\caption{(Color online) Comparison of the
experimental excitation functions of fission (capture) obtained from Refs. \cite{Mitsuoka} (stars), \cite{Keller} (solid up triangles)  and \cite{Back53} (open squares) and  of fusion obtained from Ref. \cite{Keller} (open up triangles)) for the $^{32}$S+$^{182}$W reaction with theoretical results by the DNS model for the capture (dashed line), complete fusion (thin solid line), quasifission (dot-dashed line), fast fission (dot-dot-dashed line). Theoretical values of fusion-fission  and  fusion cross sections are nearly equal because the calculated results of the total cross section for  the evaporation residues after emission of neutrons, protons and $\alpha$-particles from $^{214}$Th$^*$ compound nucleus are small (thick solid line).}
\label{CrossSec82}
\end{figure}

 The calculations were performed for the  energy range $E_{\rm c.m.}$=119.5--220.5 MeV  and initial values of the  orbital angular momentum $\ell$ =0--150$\hbar $. Due to the deformed shape of $%
^{184}$W ($\beta _{2}=0.24$ and $\beta _{4}=-0.095$) in the ground state we
included in our calculations a dependence of the excitation function of
capture, complete fusion and quasifission on the orientation angle  $\alpha _{\rm T}$ of its axial symmetry axis. The ground state shape of $^{32}$S is
spherical but the quadrupole ($2^{+}$) and octupole ($3^{-}$) collective
excitations in spherical nuclei are taken into account as amplitudes of the zero-point motion mode of surface vibration by the same way as in Ref. \cite{FazioJSot2008}. The deformation parameters of the first excited quadrupole state $\beta _{2}^{(2+)}=0.312$ (taken from Ref. \cite{RamanADND1987}) and the ones of the first excited octupole state $\beta _{3}^{(3-)}$=0.41 (taken from Ref. \cite{SpearADND1989}).
The final results of the capture and complete fusion are obtained by
averaging the contributions calculated for the different orientation angles
($\alpha _{\rm T}$=0$^{\circ }$, 15$^{\circ }$, 30$^{\circ }$, ..., 90$^{\circ }$) of the symmetry axis of the target nucleus:
\begin{equation}
\label{averfus}
\langle \sigma _{\rm fus}(E_{\rm c.m.},l)\rangle =\int_{0}^{\pi /2}\sin\alpha
_{\rm T}\sigma _{\rm fus}(E_{\rm c.m.},l;\alpha _{\rm T})d\alpha _{\rm T}.
\end{equation}
These methods were developed and used in the Refs. \cite{NasirovNPA759,NasirovPRC79,FazioJSot2008}.
 The partial cross sections of CN formation at the given energy $E_{\rm c.m.}$
are used to calculate the ER formation and fusion-fission cross sections by
the advanced statistical model \cite{ArrigoPRC1992,ArrigoPRC1994,SagJPG1998}.
 The code takes into account the competition between
evaporation of light particles (n, p, $\alpha$, and $\gamma$) and
fission processes along each step of the deexcitation cascade of
CN. The effective fission barrier for CN and intermediate excited
nuclei along the cascade are obtained taking into account the
macroscopic fission barrier, predicted by the rotating droplet
model as parameterized by Sierk \cite{SierkPRC1986}, together with
the microscopic corrections allowing for the angular momentum and
temperature fade-out of shell corrections \cite{MollerADND1988} to
the fission barrier (see Refs.
\cite{FazioEPJ2004,FazioMPL2005,FazioEPJ200475}, and reference
therein).
The cross section of ER formed at each step $x$ of the deexcitation cascade after the emission of $\nu(x)$n+$y(x)$p+$k(x) \alpha+s(x)\gamma$ particles ($\nu,y,k,s$ are numbers of neutrons, protons, $\alpha$-particles and $\gamma$-quanta) from the hot CN is calculated by the formula\cite{FazioPRC2005,FazioMPL2005,FazioEPJ2004}:
\begin{equation}
\label{ERcascade}
\sigma_{\rm ER}(E^*_{x})=\sum\limits_{Z_{l=0}}^{l_{d}
}\sigma^{l}_{(x-1)}(E^{*}_{x})W_{{\rm sur}(x-1)}(E^{*}_{x},l),
\end{equation}
where $\sigma^{l}_{(x-1)}(E^{*}_{x})$ is the partial cross section
of the intermediate nucleus formation at the $(x-1)$th step and
$W_{{\rm sur}(x-1)}(E^{*}_{x},\ell)$ is the survival probability
of the $(x-1)$th intermediate nucleus against fission along the
deexcitation cascade of CN; $E^{*}_{x}$ is an excitation energy of
the nucleus formed at the $x$th step of the deexcitation cascade.
It is clear that $\sigma^{l}_{(0)}(E^{*}_{0})=
\sigma^{l}_{\rm fus}(E^{*})$ at $E^{*}_{\rm CN}=E^{*}_{0}=E_{\rm
c.m.}+Q_{\rm gg}$, where $Q_{\rm gg}$ is energy balance of
reaction. The numbers of the being emitted neutrons, protons,
$\alpha$-particles, $\gamma$-quanta, $\nu(x)$n, $y(x)$p,
$k(x)\alpha$, and $s(x)\gamma$, respectively, are functions of the
step $x$. The emission branching ratio of these particles depends
on the excitation energy and angular momentum of the being cooled
intermediate nucleus $A=A_{\rm CN}-(\nu(x)+y(x)+4k(x))$ and
$Z=Z_{\rm CN}-(y(x)+2k(x))$ \cite{FazioEPJ2004}.
\begin{figure}[t]
\vspace*{-0.40cm}
\par
\begin{center}
\resizebox{0.80\textwidth}{!}{\includegraphics{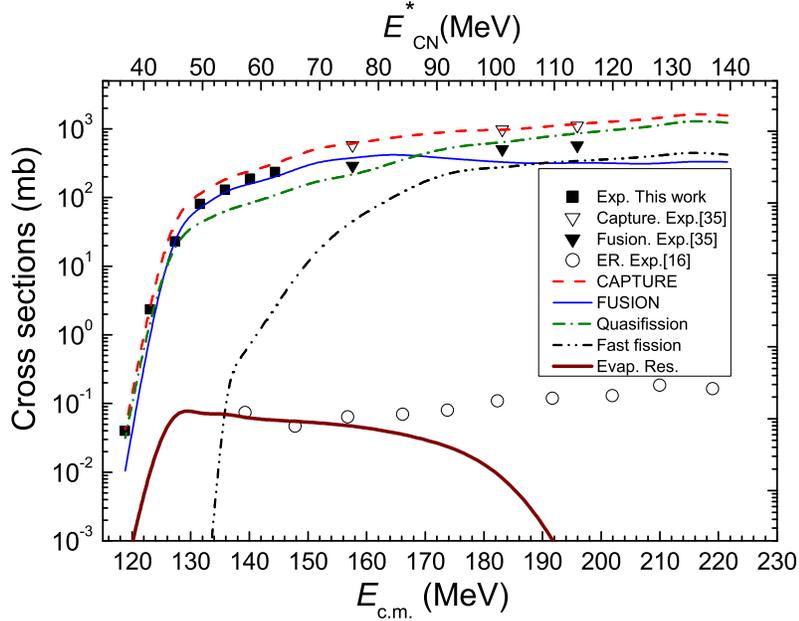}}
\vspace*{-1.6cm}
\end{center}
\caption{(Color online) Experimental excitation functions of capture obtained in this work (squares) and in Ref. \cite{DJHofmann} (open down triangles), of fusion obtained in Ref. \cite{DJHofmann} (solid down triangles),  of evaporation residues after emission of neutrons, protons and $\alpha$-particles from $^{216}$Th$^*$ compound nucleus for the $^{32}$S+$^{184}$W reaction \cite{Back60} (open circles) are compared with theoretical results by the DNS  model for the capture (dashed line), complete fusion (thin solid line), quasifission (dot-dashed line), fast fission (dot-dot-dashed line) and total evaporation residues (thick solid line). Theoretical values of fusion-fission  and  fusion cross sections are nearly equal because the total cross section for the evaporation residues after emission of neutrons, protons and $\alpha$-particles from $^{216}$Th$^*$ compound nucleus are very small.
}
\label{CrossSec84}
\end{figure}
The results of calculation of the excitation functions for the $^{32}$S+$^{184}$W and $^{32}$S+$^{182}$W reactions are presented in Figs. \ref{CrossSec84} and \ref{CrossSec82}, respectively, and they are compared with the available experimental data. In Fig. \ref{CrossSec84}, the excitation functions of capture obtained in this work (squares) and in Ref. \cite{DJHofmann} (open down triangles), of complete fusion obtained in Ref. \cite{DJHofmann} (solid down triangles),  of evaporation residues after emission of neutrons, protons and $\alpha$-particles from $^{216}$Th$^*$ compound nucleus for the $^{32}$S+$^{184}$W reaction \cite{Back60} (solid circles) with theoretical results by the DNS model for the capture (dashed line), complete fusion (thin solid line), quasifission (dot-dashed line), fast fission (dot-dot-dashed line) and total evaporation residues (thick solid line). Theoretical values of fusion-fission  and  fusion cross sections are nearly equal because the total cross section for the evaporation residues after emission of neutrons, protons and $\alpha$-particles from $^{216}$Th$^*$ compound nucleus are small.
The agreement between the experimental and theoretical capture cross sections was reached by changing the radius parameter $C_R$ entered to rescale the
nuclear radius:
\begin{equation}
R_1=C_R \sqrt{(R_{\rm p}^2\,Z_1+R_{\rm n}^2\, (A_1-Z_1))/A_1} \,,
\end{equation}
where $R_{\rm p}$ and $R_{\rm n}$ are the proton  and neutron radii, respectively, obtained from Ref. \cite{Pomorski}:
\begin{eqnarray}
R_{\rm p}&=&1.237 (1-0.157 (A-2 Z)/A-0.646/A)A^{1/3}, \\
R_{\rm n}&=&1.176 (1+0.25 (A-2 Z)/A+2.806/A)A^{1/3}.
\end{eqnarray}
The presented results are obtained at  $C_R=0.925$ for all values
of $E_{\rm c.m.}$. Using Eq.(\ref{averfus}) we calculated the
partial fusion cross sections which were used to estimate the
cross sections of ER and fusion-fission by the advanced
statistical model \cite{ArrigoPRC1992,ArrigoPRC1994}. Taking into
account the dependence of the fission barrier ($B_{\rm f}$) of the
rotating CN on its angular momentum  we found  a value of $\ell$
at which $B_{\rm f}$ disappears using the rotating finite range
model by A. J. Sierk \cite{SierkPRC1986}: $\ell_{\rm B}$=68 for $^{214}$Th 
and  $\ell_{\rm B}$=70 for $^{216}$Th. Then
we calculate the fast fission contribution for $\ell>\ell_{\rm B}$
\begin{equation}
\sigma_{\rm fast\,fission}(E_{\rm c.m.})=\sum_{\ell=\ell_{\rm B}}^{\ell=\ell_{\rm max}}(2\ell+1)\sigma_{\rm fus}(E_{\rm c.m.},\ell)
\end{equation}
where $\ell_{\rm max}$ is the maximum value of angular momentum of the DNS for the given value of $E_{\rm c.m.}$. The value of $\ell_{\rm max}$ is found by solving  the equations of motion for the radial distance and orbital angular momentum with the given values of $E_{\rm c.m.}$, $\ell_0$ and $R_{\rm max}=20$ fm.

In Fig. \ref{CrossSec84}, the ER cross sections obtained by the advanced statistical model (see Refs. \cite{ArrigoPRC1992,ArrigoPRC1994,SagJPG1998}, and
references therein) describing the full deexcitation cascade of
the $^{216}$Th$^{*}$ CN formed in the $^{32}$S+$^{184}$W
reaction are compared with the corresponding experimental data from Ref. \cite{Back60}. The theoretical excitation function (thick solid curve)
of the total evaporation residues is close to the measured data
\cite{Back60} up to $E_{\rm c.m.}\approx 160$. The dip of the theoretical curve at collision energies $E_{\rm c.m.} > 160$ MeV can be explained by production of the observed fragments in the nonequlibrium  mechanism without formation of the compound nucleus in the statistical equlibrium state. The measured data in Ref. \cite{Back60} could include the evaporation residues which are formed at incomplete fusion or multinucleon transfer reactions at $E_{\rm c.m.} > 160$ MeV. Because  the number of evaporation residues formed from the heated and rotating CN should not increase at decreasing the complete fusion cross section by the increase in collision energy $E_{\rm c.m.}$. The hindrance to complete fusion at large beam energies is connected by the dependence of the quasifission and intrinsic fusion barriers of DNS on its angular momentum. 
The decrease in complete fusion probability at large collision energies is connected with the increase in the quasifission  and  fast fission events which are presented in Figs. \ref{CrossSec84} and \ref{CrossSec82}. The ambiguity of the determination of complete fusion cross section from the measured yield of fissionlike products casts doubt on the reconstruction of complete fusion mechanism. Therefore, in the next Section, we will discuss the mechanisms causing hindrance to complete fusion.

\subsection{Two regions of strong hindrance to complete fusion}

\begin{figure}[tbp]
\vspace*{-0.5cm}
\par
\begin{center}
\resizebox{0.60\textwidth}{!}{\includegraphics{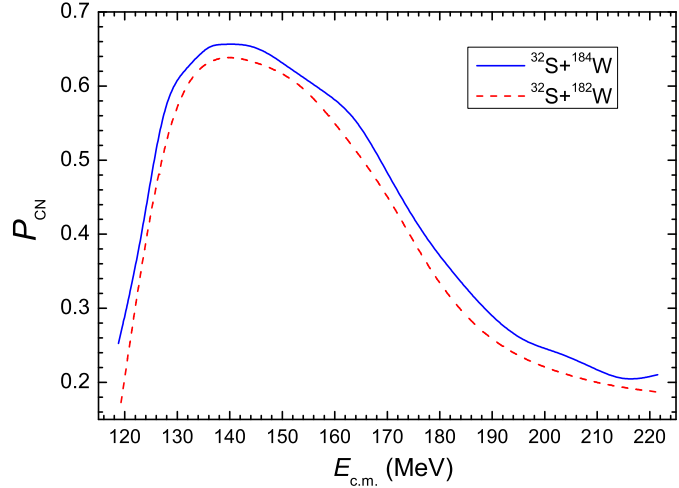}}
\vspace*{-1.35cm}
\end{center}
\caption{(Color online) Theoretical values of the fusion
probability $P_{\rm CN}$ for the $^{32}$S+$^{182}$W (dashed line) and $^{32}$S+$^{184}$W (solid line) reactions as a function of collision energy $E_{\rm c.m.}$. }
\label{PcnEcm}
\end{figure}

The experimental fission excitation function is decomposed into contributions  of fusion-fission, quasifission, and fast fission.
The intense of hindrance to complete fusion is estimated by the fusion factor $P_{\mathrm{CN}}$ which was entered in Eq. (\ref{fusion}) and it is determined by Eqs. (\ref{PcnG}) and (\ref{PcnY}).  In Fig. \ref{PcnEcm} the calculated values of $P_{\mathrm{CN}}$ as a function of the
beam energy for the $^{32}$S+$^{182}$W (dashed line) and $^{32}$S+$^{184}$W (solid line) reactions are presented. It is seen the hindrance to fusion is strong  at very small and large values of the collision energy $E_{\rm c.m.}$ for the both reactions. According to the DNS model, the use of the heavy $^{184}$W isotope was more favorable to CN formation in comparison with  using $^{182}$W.  It is explained by the fact that quasifission barrier $B_{\rm qf}$ (see Fig. \ref{PES216Th}b) for the reaction with $^{184}$W is larger than one for the reaction with  $^{182}$W.
 The yield of quasifission is dominant at the subbarrier  beam energies leading to capture of deformed nuclei only with the small orientation angle of its symmetry axis relative to the beam direction. This phenomenon was found by Hinde and his colleagues \cite{HindePCR53} in the $^{16}$O+$^{238}$U reaction where they observed the increase in the anisotropy of angular distribution of the fission fragments when beam energy decreases to the subbarrier region. Its analysis has been discussed in Ref. \cite{anisEPJA34} in connection with the explanation of the observed large angular anisotropy of the fission fragments. As an alternative suggestion for the origin of the observed anomaly, Liu \textit{et al.} in Ref. \cite{ZhangPLB353} have put forward a new version of the
preequilibrium fission model. The anomalous bump exists in the variation of
the fragment anisotropy with the incident energy for the systems such as $%
^{19}$F, $^{16}$O +$^{232}$Th. Based on these studies, a new model of $K$
pre-equilibrium fission was proposed, which can well explain the observed
anomalous anisotropy.
The authors of Ref. \cite{HindePCR53} analyzed in
detail the angular anisotropy of fragments at low energies to show the
dominant role of the quasifission in collisions of the projectile with the
target nucleus when the axial symmetry axis of the latter is oriented along
or near the beam direction. Large values of the anisotropy were obtained at
low energies and these data were assumed to be connected with quasifission
because a mononucleus or DNS formed in the near tip collisions
has an elongated shape.
\begin{figure}[tbp]
\vspace*{-2.0cm}
\par
\begin{center}
\resizebox{0.60\textwidth}{!}{\includegraphics{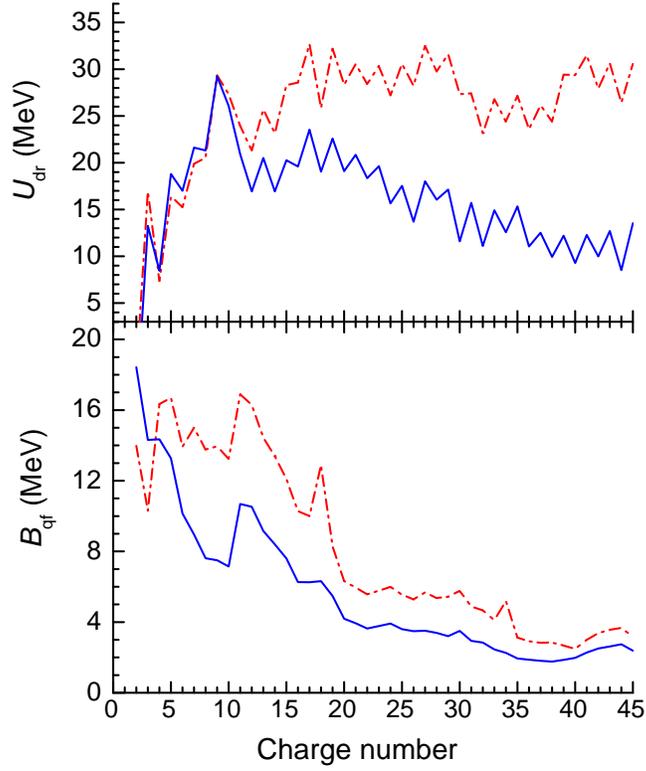}}
\vspace*{-4.0cm}
\end{center}
\caption{(Color online) Driving potential calculated for the two different values of the
orientation angles of target nucleus: $\protect\alpha _{\rm T}=15^{\circ }$
(solid line) and  $\protect\alpha_{\rm T}=45^{\circ }$ (dot-dashed line) for the $^{32}$S+$^{184}$W reaction.}
\label{UdrivBqf}
\end{figure}
This shape can be far from the one corresponding to the saddle
point. The driving potential  used in the DNS model depends on the
shape of the DNS which is formed in collisions with the
different orientation angles of the axial symmetry axis of
deformed target. The small values of the fusion probability
$P_{\mathrm{CN}}$ at lowest energies in the $^{32}$S+$^{182,184}$W
reactions are explained by the large values of $B^*_{\rm fus}$ for 
DNS formed at collisions of projectile with
target oriented close to beam direction, {\it i.e.} when
$\alpha_{\rm T}$ is small. To demonstrate this conclusion  we present in
Fig. \ref{UdrivBqf} the driving potential (upper panel) and
quasifission barrier (lower panel) for the decay of DNS as a function of the orientation angles of the target
nucleus. It is seen from the upper panel of Fig. \ref{UdrivBqf}
that the value of driving potential corresponding to the
projectile charge number $Z$=16 for the small orientation angle
15$^{\circ}$ (solid line) of the target is lower than that
corresponding to the larger orientation angle 45$^{\circ}$ (dashed
line). The fusion probability is larger if a value of the driving
potential $U_{\mathrm{dr}}(Z_{\mathrm{proj}})$
corresponding to the initial charge number of the light fragment $Z_{\mathrm{%
L}}$ of the DNS is at the same level with the maximum value at $Z$=9 along the way to complete fusion or as possible higher than this maximum value. In these cases,  all nucleons of light nucleus transfer easy into heavy nucleus: $Z_{\newline
\rm L}\longrightarrow 0$ and $Z_{\mathrm{H}}\longrightarrow Z_{\mathrm{CN}}$.
\begin{figure}[tbp]
\par
\begin{center}
\resizebox{0.60\textwidth}{!}{\includegraphics{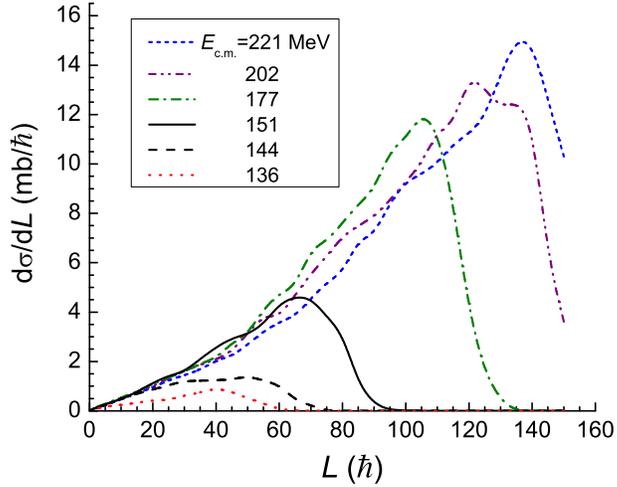}}
\vspace*{-1.0cm}
\end{center}
\caption{(Color online) The partial quasifission excitation function  calculated at different values of the collision energy  $E_{\mathrm{c.m.}}$ for the $^{32}$S+$%
^{184}$W reaction.}
\label{QuasifissEL}
\end{figure}

The lowering of the $P_{\rm CN}$ values in Fig. \ref{PcnEcm} at collision energies $E_{\mathrm{c.m.}}>145$ MeV is explained by decreasing the quasifission barrier $B_{\rm qf}$ as a function of the orbital angular momentum $\ell$. Because the depth of the potential well being $B_{\rm qf}$ for a given charge asymmetry decreases due to the increase
in the rotational energy $E_{\mathrm{rot}}$ of the DNS 
(for details see \cite{FazioPRC2005}). At the same time the intrinsic fusion barrier $B_{\mathrm{fus}}^{\ast }$ increases by increasing $\ell$. At small collision energies   only small values of $\ell$ are populated because a capture does not occur if the initial energy of projectile is not enough to overcome the
Coulomb barrier of the entrance channel.

\begin{figure}[tbp]
\par
\begin{center}
\resizebox{0.60\textwidth}{!}{\includegraphics{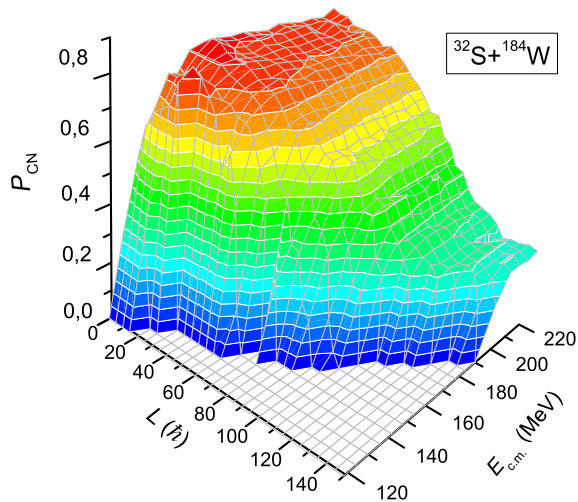}}
\vspace*{-1.0cm}
\end{center}
\caption{(Color online) The presentation of the fusion probability $P_{\rm CN}$ for the $^{32}$S+$^{184}$W reaction as a function of collision energy $E_{\mathrm{c.m.}}$ and initial angular momentum $L$.}
\label{Pcn3D}
\end{figure}

It is seen from Fig. \ref{QuasifissEL} that the values $\ell>70$ are populated at collision energies $E_{\mathrm{c.m.}}> 144 $ MeV. Therefore, we conclude that the contribution of quasifission and fast fission becomes dominant at $E_{\mathrm{c.m.}}> 144 $ MeV and it has an effect on the anisotropy of angular distribution which increases by increasing the collision energy.  In Fig. \ref{Pcn3D}, the appearance of the quasifission as a hindrance to fusion is demonstrated as a function of the collision energy and orbital angular momentum for the $^{32}$S+$^{184}$W reaction. One can see that  maximum value of $P_{\rm CN}$ (favorable condition to complete fusion) is reached at their middle values of $E_{\mathrm{c.m.}}$=135-140 MeV and at $\ell$=20--25.
This is connected with the fact that at these energies all orientation angles $\alpha_{\rm T}$ can contribute to the formation of compound nucleus because the Coulomb barrier for large orientation angles may be overcame. As we know, in collisions with large $\alpha_{\rm T}$ the fusion probability is large. The decrease of $P_{\rm CN}$ at  larger energies is explained by the decrease in the quasifission barrier $B_{qf}$ by increasing $\ell$ in collisions with all orientation angles.

In synthesis of superheavy elements by using actinide nuclei as a
target the beam energy have to be larger  enough than the well
known Bass barrier to have possibility to include contribution of
the large orientation angles. Therefore, a further increase in the
beam energy leads not only to a decrease in the survival
probability of the  hot CN but also to the strong increasing the 
quasifission events. The measured anisotropy of the fragments ascribed as
fission products confirmed this conclusion.

\subsection{Anisotropy of fission fragment angular distribution and variance $K_0^2$ of $K$ distribution}

 To clarify the role of quasifission fragments in the observed anisotropy $A\rm _{exp}$ of the fission fragment angular distribution, we calculated the anisotropies of the angular distributions  of the quasifission and fusion-fission fragments. We used the  approximated expression to calculate the anisotropy  suggested by Halpern and Strutinski in Ref. \cite{Ha1958} and Griffin in Ref. \cite{GriPR1959}:
\begin{equation}
\label{anisJeff}
\mathcal{A}\approx{1+}{\frac{\langle{l^2}\rangle_{i}{\hbar}^2}{4{\Im_{\rm eff}}^{(i)}T_{i}}}, \,{\rm where} \,\,\frac{1}{\Im_{\rm eff}}=\frac{1}{\Im_{\|}}-\frac{1}{\Im_{\bot}}.
\end{equation}
Here $\Im_{\rm eff}^{(i)}$ is the effective moment of inertia for the CN on the saddle point  $i={\rm CN}$ or it is the effective moment of inertia
for the DNS on the quasifission barrier $i={\rm DNS}$.
In the last case,  we calculate $\Im_{\rm eff}$ for the DNS taking into account the possibility of different orientation angles of its constituent nuclei (see Appendix A of Ref.\cite{anisEPJA34}), assuming that after capture the mutual orientations of the DNS nuclei do not change sufficiently.
 $\Im_{\|}$ and $\Im_{\bot}$ are the moments of inertia
around the symmetry axis and a perpendicular axis, respectively.
\begin{figure}[tbp]
\includegraphics[width=12.cm,angle=0.]{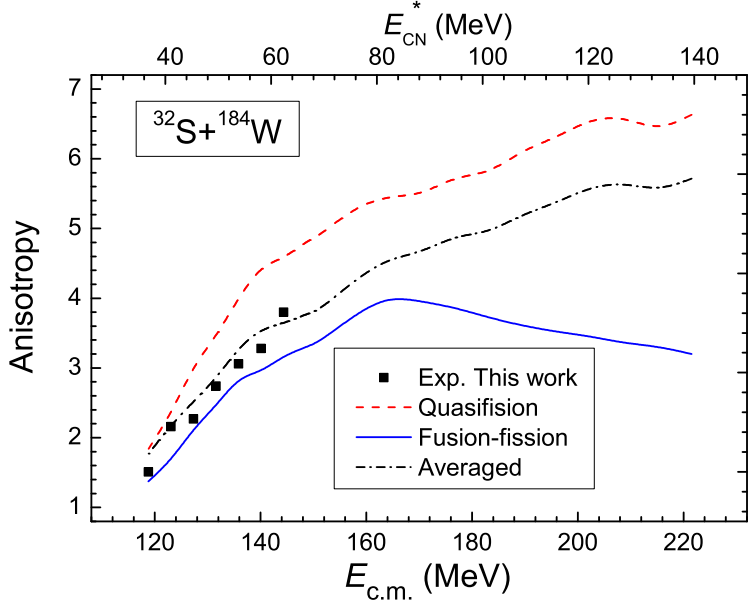}
\vspace*{-0.6 cm}
\caption{(Color online) Comparison of the anisotropy measured (circles) in this work for the
$^{32}$S+$^{184}$W reaction with the theoretical results for the
anisotropy of the quasifission (dashed line) and fusion-fission
(solid line) fragments as function of the center-of-mass energy
(bottom axis) and excitation energy of CN (top axis). The averaged
values of theoretical results are presented by dot-dashed line.}
\label{FigAnis}
\end{figure}

Their values for $\Im_{\rm eff}^{(\rm CN)}$ are determined in the framework of the
rotating finite range model by Sierk \cite{SierkPRC1986}. The
temperature of CN on the saddle point is found by the expression:
\begin{equation}
T_{\rm CN}=\left[\frac{E_{\rm c.m.}+Q_{\rm gg}-
B_{\rm f}(\ell)-E_n}{A_{\rm CN}/12}\right]^{1/2},
\end{equation}
where $B_{\rm f}(\ell)$ is the fission barrier height. $B_{\rm
f}(\ell)$ is calculated in terms of the rotating liquid drop model
by Sierk \cite{SierkPRC1986}. $E_{\rm n}$ is the energy carried away
by the pre-saddle fission neutrons. The temperature of DNS on the
quasifission barrier is determined by the expression:
\begin{equation}
T_{\rm DNS}=\left[\frac{(E^*_{\rm DNS}-B_{\rm qf})}{A_{\rm CN}/12}\right]^{1/2}.
\end{equation}
An important physical quantity in the formula (\ref{anisJeff}) is the variance $K^{2}_{0}$ of the Gaussian distribution of the $K$ projection :
\begin{equation}
K^{2}_{0}=\frac{\Im_{\rm eff}T_{\rm saddle}}{\hbar^2}.
\end{equation}
The experimental values of $K^2_{0}$ are used to fit the angular distribution of fission fragments by formula (\ref{Wangl}) (see Ref.\cite{Hui177,Hinde}).
\begin{figure}[tbp]
\vspace*{0.5cm}
\includegraphics[width=10.cm,angle=0.]{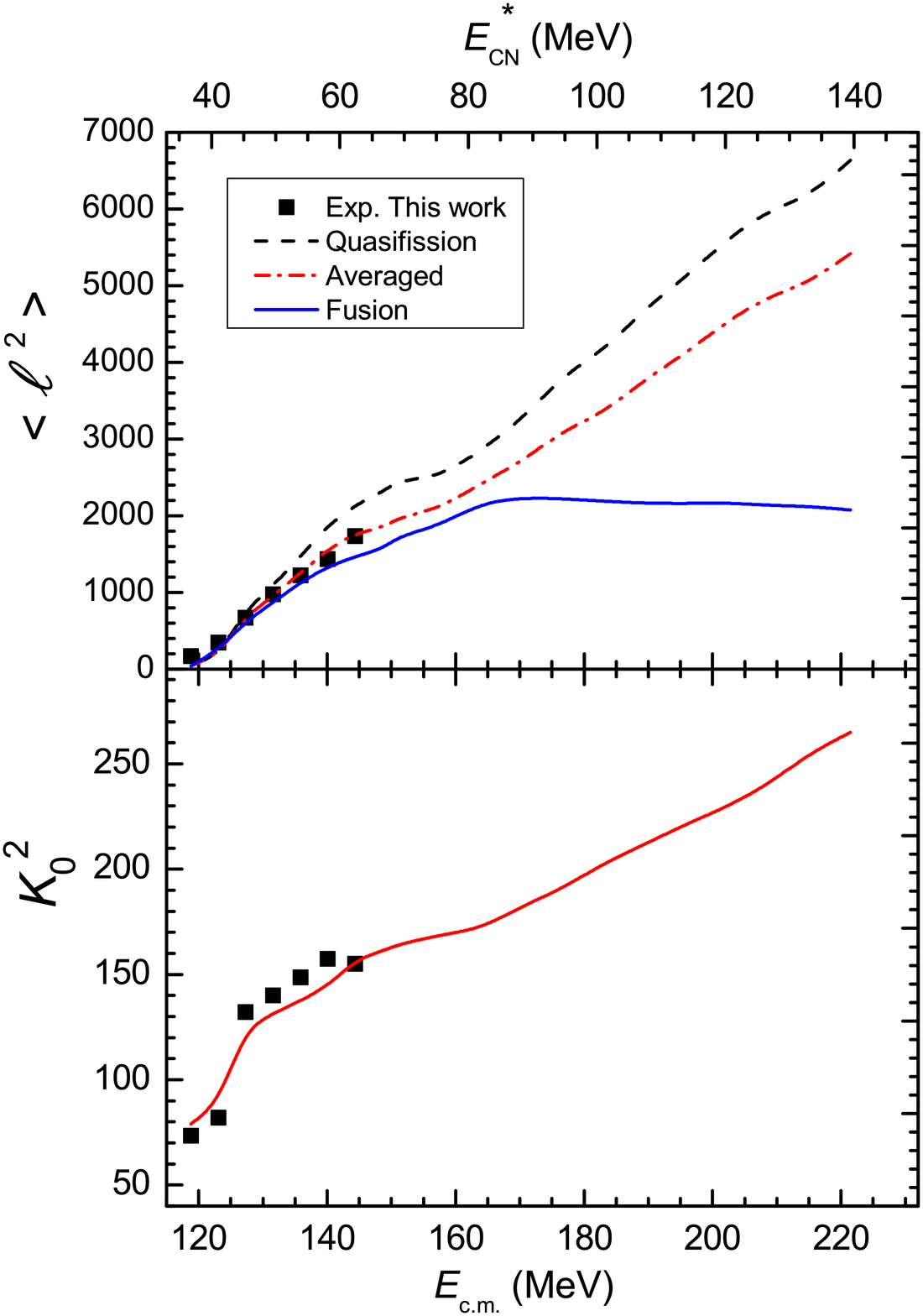}
\vspace*{-2.0cm}
\caption{(Color online) Mean square value of the angular momentum (upper panel) versus the collision energy $E_{\rm c.m.}$ for the  $^{32}$S+$^{184}$W reaction. The experimental data (solid squares) are shown in comparison with the results of DNS model which are obtained by averaging $\ell^2$ by partial cross sections of the quasifission  (dashed line) and complete fusion (solid line) events. In the lower panel the  experimental data of this work (solid squares) for  $K_0^2$ are compared with our theoretical results.}
\label{FigL2}
\end{figure}

In Fig. \ref{FigAnis} we compare the anisotropy measured (squares)
in this work with the theoretical results for the anisotropy of
the quasifission (dashed line) and fusion-fission (solid line)
fragments as a function of the center-of-mass energy (bottom axis)
and excitation energy of the CN (top axis). The averaged
theoretical anisotropy  over the contributions of both mechanisms
are presented by the dot-dashed line. It is seen that the averaged
values of anisotropy  $\mathcal{A}$ are closer to the experimental data. Consequently we confirm that the measured cross section of
the fission fragment formation and their angular distribution are
results of mixing of the quasifission and fusion-fission products.
The dip of the solid curve corresponding to the contribution of the fusion-fission fragments is caused by the increasing  the effective temperature of CN with the angular momentum $\ell < \ell_f=70$. 
 
In the upper panel of Fig.\ref{FigL2}, we compare  experimental
and theoretical values of mean square values of angular momentum.
The experimental $\langle {\ell^{2}}\rangle$ values are obtained from
the measured anisotropy $\mathcal{A}_{\rm exp}$ and $K_0^2$ values used to fit
measured angular distributions presented in Fig. \ref{fig_2}. The
theoretical values for fusion-fission and quasifission fragments
are calculated by averaging  $\ell^2$ using the partial cross
sections of the quasifission  (dashed line) and complete fusion
(solid line) events. The experimental data are well described with
the averaged values of  $\ell^{2}$\ between the complete fusion and
quasifission cross sections:
\begin{equation}
\label{avL2}
\langle {\ell^{2}}\rangle=\frac{(\sigma_{\rm fus}\langle {\ell^{2}}\rangle_{\rm fus}+\sigma_{\rm qf}\langle {\ell^{2}}\rangle_{\rm qf})}{\sigma_{\rm fus}+\sigma_{\rm qf}}.
\end{equation}
In the lower panel of  Fig.\ref{FigL2}, the experimental data of this work (solid squares)  for $K_0^2$
are compared with the theoretical values obtained from the description of $\mathcal{A}_{\rm exp}$ (dot-dashed line in Fig.\ref{FigAnis}) and  $\langle {\ell^{2}}\rangle$ (dot-dashed line in Fig.\ref{FigL2}) extracted from the experimental data of the angular distribution of fission fragments. This comparison shows again the dominancy of the quasifission fragments into measured data.

\begin{figure}[tbp]
\includegraphics[width=15.cm,angle=0.]{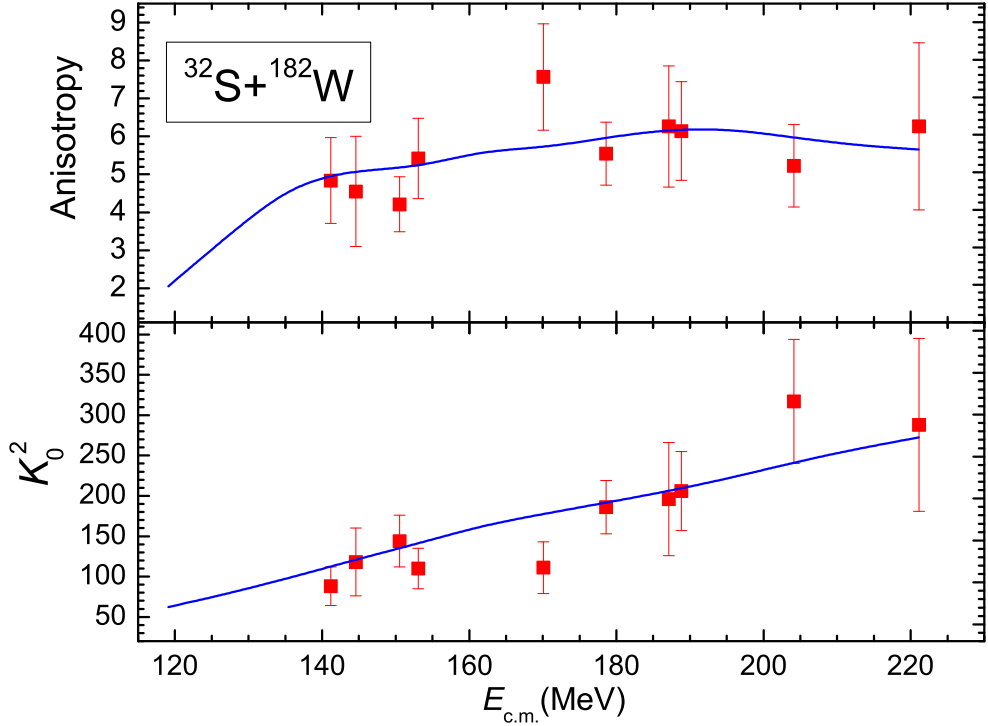}
\caption{(Color online) Comparison of the experimental data from Ref. \cite{Keller}  for   $K_0^2$ (solid squares) and anisotropy $\mathcal{A}$  with results of the DNS model (solid line).}
\label{K02S182W}
\end{figure}

In Fig. \ref{K02S182W} we compare the results of our calculation for $K_0^2$ with the experimental data presented in Ref. \cite{Back53} for the $^{32}$S+$^{182}$W reaction as a function of $E_{\rm c.m.}$.  The good description of the experimental data of the capture excitation function for both reactions and quantities $K_0^2$, $\mathcal{A}_{\rm exp}$ and $<\ell^2>$  which characterize angular distribution of the fission products by the partial capture  and  fusion cross sections proves the rationality of the theoretical method based on the DNS concept to analysis the fusion-fission,  quasifission and fast fission mechanisms at the considered range of beam energy. We should stress that the partial fusion cross sections were used to calculate excitation function of the total evaporation residues by the advanced statistical model \cite{ArrigoPRC1992,ArrigoPRC1994,SagJPG1998}. 

\section{CONCLUSION}
\label{sec:5}

The fission angular distributions for the $^{32}$S+$^{184}$W reaction have been measured at center-of-mass energies of 118.8--144.4 MeV.
The experimental cross sections of fissionlike products, the anisotropy
$\mathcal{A}_{\rm exp}$, $K_0^2$ and $\langle{\ell^{2}}\rangle$ values were obtained.  The experimental data of this work, of Refs. \cite{DJHofmann} (capture exciation function) and  \cite{Back60} (excitation function of the total evaporation residues) for this reaction and the same kind of experimental data for the $^{32}$S+$^{182}$W reaction obtained from Refs. \cite{Mitsuoka,Keller,Back53} (capture and fusion) and  \cite{Back53,Keller} ($K_0^2$ and $\ell^2$) were described by the DNS model and advanced statistical model \cite{ArrigoPRC1992,ArrigoPRC1994,SagJPG1998}.
The measured fission excitation function was described as the capture excitation function containing quasifission and fast fission products together with fusion-fission products.  The partial capture cross sections were calculated for the different orientation angles of the symmetry axis of the target nuclei $^{182}$W and $^{184}$W. 
The quadrupole ($2^{+}$) and octupole ($3^{-}$) collective excitations in $^{32}$S are taken into account as amplitudes of the zero-point motion mode of surface vibration by the same way as in Ref.\cite{FazioJSot2008}.
The total evaporation residue and fusion-fission
excitation functions are  calculated in the framework of the
advanced statistical model using the partial fusion cross sections obtained in this work for both considered reactions. The dip of the theoretical curve from experimental data \cite{Back60} for the $^{32}$S+$^{184}$W reaction  at high excitation energies $E_{\rm c.m.} > 160$ MeV is caused by the fact that statistical model can not reproduce the cross section of formation of reaction products by the nonequlibrium  mechanism without formation of the compound nucleus in the statistical equlibrium state. The ER events at $E_{\rm c.m.} > 160$ seem to be connected with incomplete fusion  or multinucleon transfer reactions.  Because  the number of evaporation residues formed from the heated and rotating CN should not increase at decreasing of the complete fusion cross section by the increase in the collision energy $E_{\rm c.m.}$. The decrease of complete fusion probability at large collision energies is connected with the increase of quasifission  and  fast fission.
 An increase in the quasifission contribution at large beam energies is connected with the angular momentum dependence of the quasifission $B_{\rm qf}$
and intrinsic fusion $B_{\rm fus}^{\ast }$ barriers:  at large
angular momentum of the DNS  $B_{\rm qf}$ decreases and
$B_{\rm fus}^{\ast }$ increases. The small quasifission barrier
decreases the lifetime of DNS decreasing its possibility  to be transformed into a CN \cite{NasirovNPA759, FazioPRC2005}. The  fusion-fission cross section is larger than the quasifission cross section in the energy range $130 <E_{\rm c.m.} < 170$ MeV where we have $P_{\rm CN}>0.5$ for both reactions. The contribution of the fast fission becomes comparable  with the fusion-fission and quasifission products at about  $E_{\rm c.m.} >$ 175 MeV. The experimental data presented and analysed in this work are the smooth continuation to the lower energies of the data presented in previous published papers here cited. The theoretical descriptions of the experimental capture excitation function for both reactions and quantities $K_0^2$,  $<\ell^2>$ and  $\mathcal{A}_{\rm exp}$ which characterize angular distribution of the fission products were performed by the same partial capture cross sections at the considered range of beam energy. 
We conclude that the effects of competition between fusion and quasifission in the reaction play an important role in the dynamics process.

\textbf{ACKNOWLEDGMENTS}

This work was supported by the Major State Basic Research Development
Program under Grant No. G2000077400 and the National Natural Science
Foundation of China under Grant Nos. 10375095, 10735100, and 10811120019.
A. K. Nasirov is grateful to the Istituto Nazionale  di Fisica Nucleare and
Department of Physics of the University of Messina and Russian Foundation for Basic Research for the support of collaboration between the Dubna and Messina groups.

\end{document}